# A Novel Approach to Real-Time Short-Term Traffic Prediction based on Distributed Fiber-Optic Sensing and Data Assimilation with a Stochastic Cell-Automata Model

**Yoshiyuki Yajima**
NEC Corporation
Email: yoshiyuki-yajima@nec.com

**Hemant Prasad**
NEC Corporation

**Daisuke Ikefuji**
NEC Corporation

**Takemasa Suzuki**
NEC Corporation

**Shin Tominaga**
NEC Corporation

**Hitoshi Sakurai**
NEC Corporation

**Manabu Otani**
Central Nippon Expressway Company Limited

Word Count: 7144 words + 1 table (250 words per table) = 7394 words

*Y. Yajima, H. Prasad, D. Ikefuji, T. Suzuki, S. Tominaga, H. Sakurai, and M. Otani*

**ABSTRACT**
This paper demonstrates real-time short-term traffic flow prediction through distributed fiber-optic sensing (DFOS) and data assimilation with a stochastic cell-automata-based traffic model. Traffic congestion on expressways is a severe issue. To alleviate its negative impacts, it is necessary to optimize traffic flow prior to becoming serious congestion. For this purpose, real-time short-term traffic flow prediction is promising. However, conventional traffic monitoring apparatus used in prediction methods faces a technical issue due to the sparsity in traffic flow data. To overcome the issue for realizing real-time traffic prediction, this paper employs DFOS, which enables to obtain spatially continuous and real-time traffic flow data along the road without dead zones. Using mean velocities derived from DFOS data as a feature extraction, this paper proposes a real-time data assimilation method for the short-term prediction. As the theoretical model, the stochastic Nishinari-Fukui-Schadschneider model is adopted. Future traffic flow is simulated with the optimal values of model parameters estimated from observed mean velocities and the initial condition estimated as the latest microscopic traffic state. This concept is validated using two congestion scenarios obtained in Japanese expressways. The results show that the mean absolute error of the predicted mean velocities is 10–15 km/h in the prediction horizon of 30 minutes. Furthermore, the prediction error in congestion length and travel time decreases by 40–84% depending on congestion scenarios when compared with conventional methods with traffic counters. This paper concludes that real-time data assimilation using DFOS enables an accurate short-term traffic prediction.

Y. Yajima, H. Prasad, D. Ikefuji, T. Suzuki, S. Tominaga, H. Sakurai, and M. Otani

**INTRODUCTION**

Traffic congestion on expressways is a serious issue in terms of economic loss, environmental burden like air pollution, and increasing risks of accidents. They decrease service quality of road managers. To relieve such negative impacts, optimization of traffic flow is a promising solution. It enables prior alert of congestion and accidents to drivers, providing accurate prediction of travel time and congestion length, efficient patrol and the jam-absorption driving at the upstream (*1*, *2*), and decision-making support of traffic control by dynamic pricing (*3*). These actions enhance the management abilities of road managers and operators. To realize the optimization of traffic flow, developing real-time and short-term prediction is required. This makes it possible to understand when and where congestion will occur, how long congestion will extend, and what mean velocity will be during congestion beforehand under the current situation.

Short-term traffic prediction methodologies have been proposed. Recent methodologies are mainly based on machine learning (e.g., *4–7*). This approach extracts important features from extensive historical data and predicts future traffic flow according to similar patterns. Therefore, it attains beneficial performance in ordinal traffic while the accuracy decreases in the case of rare and sudden abnormal events. In addition, the statistical and time-series modeling (e.g., *4*, *8–10*) based approach have been well-established. The combination of these two is also proposed (*11*, *12*).

However, conventional traffic sensors used in previous studies involve a fundamental technical issue. Conventional apparatus is traffic counters such as surveillance cameras, loop inductors, and supersonic sensors. These point sensors have a limited observable area of a few meters–tens of a few meters. Although this problem can be solved if the sensors are installed at high density, it is not practical in terms of maintenance cost. Although methodologies to indirectly estimate traffic flow in the dead zones have been proposed (*13–16*), uncertainties will increase rather than the direct observation. Besides, probe data obtained from an on-board sensor has been recently emerged. Nevertheless, the number of samples is inadequate because the vehicles with the on-board sensor is still uncommon. Moreover, there is a time-lag to upload data at the uplink station, which reduces the real-time performance. It becomes significant when congestion occurs since the arrival time of a probe vehicle delays. The sparse traffic data makes prediction more challenging.

To overcome the issue of sparse traffic data, the application of the distributed fiber-optic sensing (DFOS) is proposed (*17–20*). DFOS detects and localizes vibration sources along the optical fiber based on Rayleigh backscattering. It is cost-efficient wide-area traffic monitoring solutions with already installed fiber infrastructures. **Figure 1** shows a schematic view of traffic monitoring using DFOS where a sensing apparatus is connected to an optical fiber cable laid along a road. Vehicle vibrations along the road can be localized by the sensing apparatus and they are seen as vehicle trajectories on the spatiotemporal plane. Namely, the fiber cable is used as a one-dimensional continuous sensor array for traffic monitoring in real-time.

Moreover, the prior art (*21*) develops a traffic monitoring system using DFOS for practical use. The sensitivity in DFOS decreases if the distance of vehicles from the cable increases due to attenuation, and consequently, trajectories are missed. The prior art (*21*) resolves the practical issue by developing an image segmentation-based trajectory recognition approach as a feature extraction to estimate mean velocity. The error in estimated mean velocity by this system compared with conventional traffic counters is ~4% for real expressway data. This technology has been implemented and commercialized by expressway operators in Japan.

This paper demonstrates a real-time short-term traffic flow prediction method that utilizes the real-time continuous traffic monitoring capabilities of DFOS. We develop a novel model-driven prediction method using mean velocities measured from DFOS through sequential data assimilation. Here, parameters in a theoretical traffic model are calibrated from observed mean velocities and future traffic flow is simulated to predict the traffic state. This approach enables reliable prediction of various traffic states even in abnormal scenarios because the simulation model is continuously updated for the current state obtained from real-time data. The fundamental framework is proposed in our previous work (*22*) based on a numerical experiment. The objective of this paper is to validate our prediction method for congestion scenarios in real traffic data measured on two expressways in Japan.

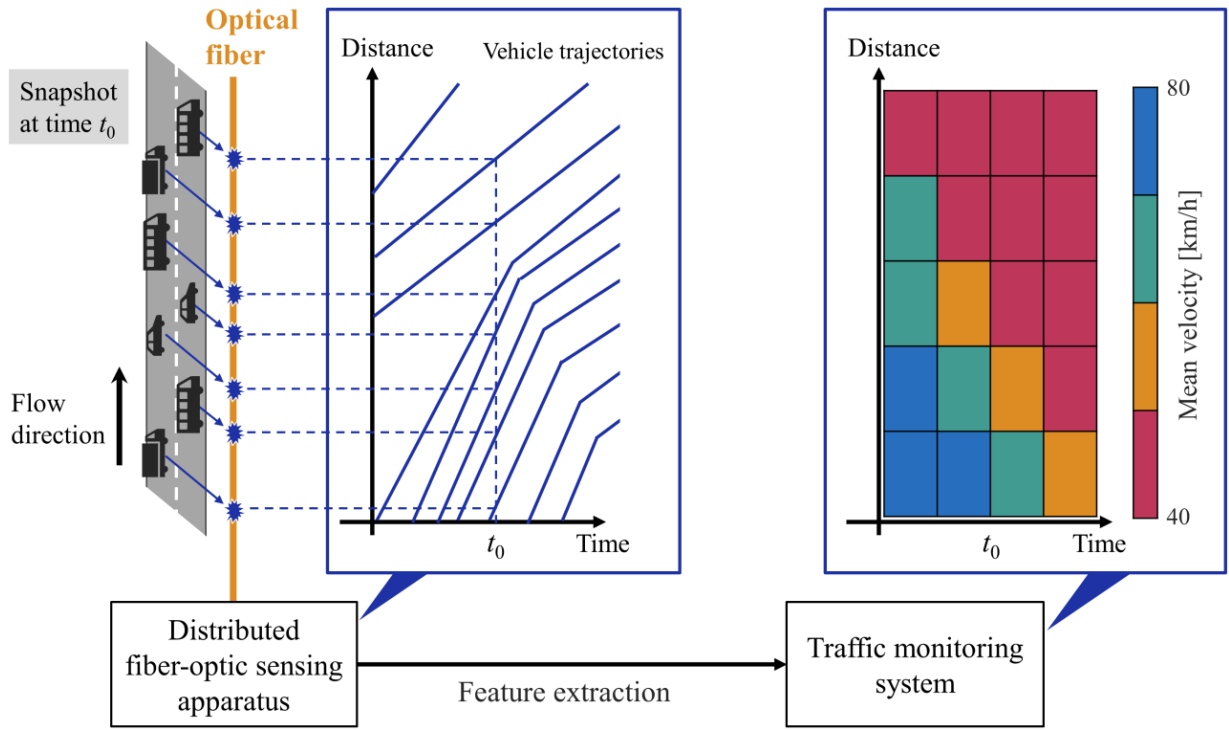

**Figure 1 Schematic view of traffic monitoring using DFOS.**

**DATA ASSIMILATION AND PREDICTION METHODS**

For traffic simulation, this paper adopts a cellular automata (CA) model. CA models divide the road into cells and consider that one vehicle occupies one cell. The models impose simple rules of vehicle behavior in each time-step. Thus, it is a self-organization system; the simple rules generate complex and realistic traffic flow in the simulation. Moreover, they are simple processes that have appropriate computational capabilities with recently improved machine power for traffic modeling and simulation.

This paper employs the stochastic Nishinari-Fukui-Schadschneider (S-NFS) model (*23*). The model considers stochastic processes of (i) involuntary deacceleration called the "random braking effect" (*24*), (ii) drivers' recognition delay or vehicle inertia called the "slow-to-start effect" (*25–27*), and (iii) drivers' anticipation called the "quick start effect" (*27–29*). These three processes are stochastically considered with probability $p$, $q$, and $r$, respectively. This paper refers to these probabilities as "model parameters". By adopting a specific value for model parameters, other traffic models are derived. Thus, it is one of the most comprehensive traffic models and adopted in this work for this reason.

**Figure 2** shows the outline of our traffic prediction method. In step 1, simulation data sets are generated by the S-NFS model adopting many model-parameter sets to emulate observed mean velocities by DFOS for the last tens of a few minutes. In step 2, the optimal values of the S-NFS model parameters that reproduce observed mean velocities are estimated using simulation data sets and the observed result. In step 3, the latest microscopic traffic state is estimated from the latest mean velocities. In step 4, using the estimated microscopic traffic state as the initial condition and optimal values of the model parameters, future traffic flow is simulated. This method assumes that the optimal values of model parameters last until the prediction horizon. Therefore, the prediction result is valid as long as the optimal values are not changed.

*Y. Yajima, H. Prasad, D. Ikefuji, T. Suzuki, S. Tominaga, H. Sakurai, and M. Otani*

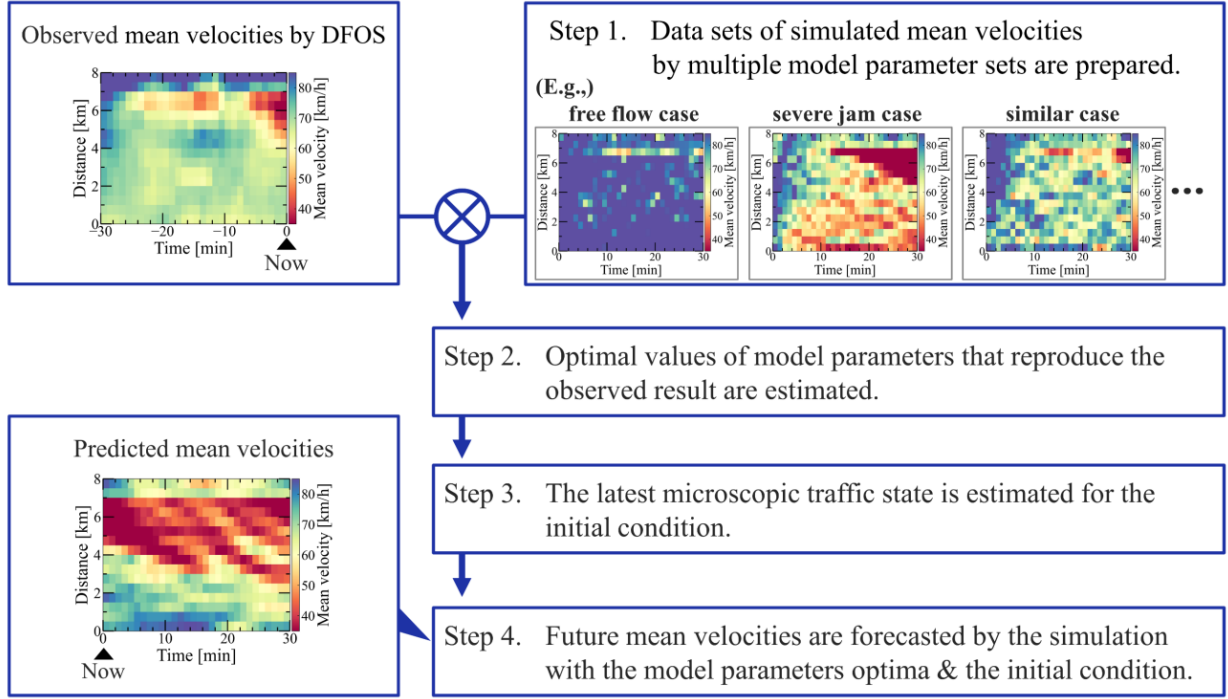

**Figure 2 Outline of our traffic prediction method using DFOS and data assimilation.**

**Estimation of Microscopic Traffic State from Mean Velocity**
Since this paper adopts a CA-based traffic model, it is necessary to estimate a microscopic traffic state such as positions and velocities each vehicle to set the appropriate initial condition of the simulations in **Figure 2** step 1 and 3. For instance, many slow vehicles should be placed where low velocities are observed in a congestion scenario. Thus, traffic flow, even in sudden abnormal congestion, can be reflected in the simulation and therefore, a reliable traffic simulation and accurate prediction can be attained.

**Figure 3** shows the estimation methodology of the microscopic state based on observed mean velocities in each road segment and the mean density–mean velocity relation, the $k$–$v$ diagram. Its principle is proposed by our previous work (*22*). The process firstly derives the mean density of vehicles in the segment and it is converted into the total vehicle numbers. Secondly, velocities of each vehicle are estimated from the total vehicle number and the mean velocity under assumptions.

The detailed methodology is as follows. The density of vehicles in a segment is estimated from observed mean velocity and the $k$–$v$ diagram. This paper adopts the Underwood model (*30*) described by **Equation 1**,

$$v = v_f \exp\left(-\frac{k}{k_m}\right), \qquad (1)$$

where $v$ is mean velocity, $k$ is mean density, $v_f$ and $k_m$ is fitting parameters. When both mean velocities and densities are obtained, $v_f$ and $k_m$ are derived from the least-squares fitting. If not, they are set as 120 km/h and 55 veh/km, respectively. These parameters reproduce a typical relation of traffic flow in Japanese expressways (*22*, *31*). From derived mean density, the total number of vehicles in the segment is estimated. If the number of lanes is larger than one, the total vehicle number on each lane is estimated by assuming the occupancy of each lane.

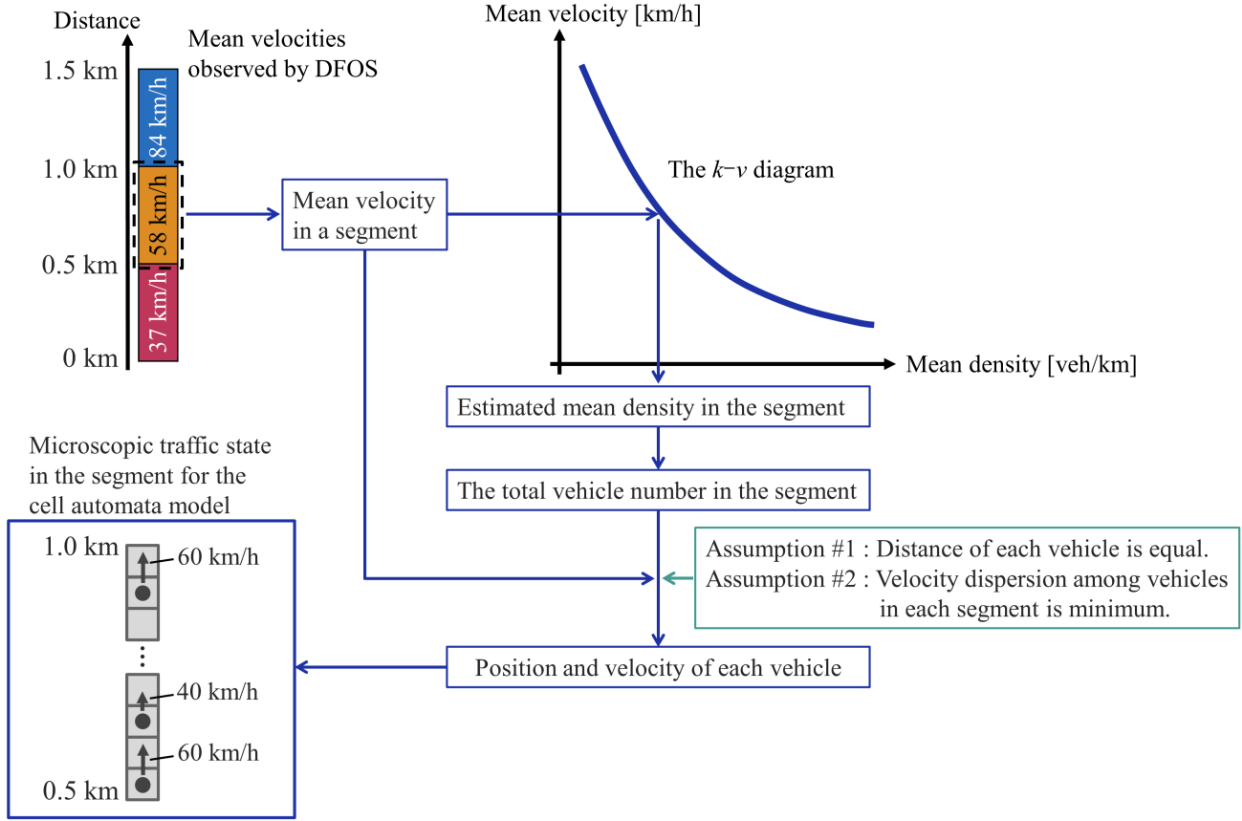

**Figure 3 Outline of microscopic traffic state estimation from observed mean velocity for CA. The figure is in the case considers the velocity resolution of 20 km/s in the CA model.**

In the next step, the positions and velocities of each vehicle are estimated. The positions can be estimated under the assumption that the distance between vehicles is equal. The velocities of each vehicle are estimated as follows. Here, we assume that the velocity dispersion in each road segment is minimal because the probability of traffic accidents is very low. For instance, suppose that the observed mean velocity is 58 km/h in a 500-m segment and the velocity resolution of the CA model is 20 km/h as shown in **Figure 2**. In this case, it is assumed that vehicles driving at only 60 and 40 km/h exist. By considering that the harmonic mean of each vehicle velocity is equal to the observed mean velocity and the assumption above, the number of vehicles driving at the lower velocity (e.g., 40 km/h in the case above) in the segment, $N_l$, can be derived from **Equation 2**,

$$N_l = \left\lfloor 0.5 + \frac{N_{\text{tot}} v_l (v_l + dv - v_s)}{v_s dv} \right\rfloor, \tag{2}$$

where $N_{\text{tot}}$ is the total number of vehicles in the segment, $v_l$ is the lower velocity, $dv$ is the velocity resolution of the CA model, $v_s$ is the mean velocity, and $\lfloor . \rfloor$ is the floor function. The number of vehicles driving at higher velocity is also derived as $N_{\text{tot}} - N_l$. Then, the velocities and positions of each vehicle are randomly combined. Processes so far are applied for all road segments where mean velocity is measured. In this way, the microscopic traffic state for the CA model considering the observed results is estimated.

*Y. Yajima, H. Prasad, D. Ikefuji, T. Suzuki, S. Tominaga, H. Sakurai, and M. Otani*

**Preparing Simulation Data Sets which Emulate Observed Traffic Flow**
As shown in **Figure 2** step 1, simulation data sets which emulate observed data are generated. To do this, the road model in the simulation is firstly developed considering the speed limit and the bottleneck position. The position of the bottleneck is identified from past observed traffic data and design drawings of the road construction. Traffic inflow rate is also used to emulate observed traffic flow. It is based on observed data of the traffic counter close to the position corresponding to the origin of the road model in the simulation. The traffic counter measures the number of vehicles per minute and the instant velocity. These data are input to the simulation system as a boundary condition.

To estimate the optimal model parameters, traffic simulation adopted many sets of the model parameters are carried out. How to estimate the optimal model parameters is described in the next subsection. Simulation time is determined according to the time-window size for the model parameter estimation. For instance, when the optimal model parameters are estimated from observed data for the latest 30 minutes, the simulation time is also set as the same time. Simulated mean velocities are derived based on the Edie method (*32*).

As the initial condition of the simulation, the microscopic traffic state estimated from the observed mean velocities in the past is adopted. The initial condition is updated every 15 minutes. That is, when the optimal model parameters are started to estimate from observed mean velocities in $t = 0$–30 minutes, the initial condition of $t = 0$–30, 1–31, …, 14–44 minutes are set as the microscopic traffic state at $t = 0$ minute. From $t = 15$–45 minutes, the initial condition is the microscopic state at the $t = 15$ minutes.

**Optimal Value Estimation of Model Parameters with the Particle Filter**
As shown in **Figure 2** step 2, the optimal values of model parameters that reproduce observed mean velocity in the theoretical model is estimated with the particle filter. The particle filter is a generalized state-space modeling method and applicable for non-Gaussian distributions and nonliner systems. Thus, it is widely used in many domains such as time-series analysis and system identification. The particle filter sequentially computes weights which are defined by the likelihood, i.e., similarity between the observed and simulated result. Finally, posterior probability distributions (PPDs) of model parameters are obtained. Optimal values are estimated from these PPDs. **Figure 4** shows the schematic view of the process. The detailed processes are as follows.

At the start time of the estimation $t_0$, $N$ sets of model parameters $\boldsymbol{\theta}_n$ ($n = 1, 2, \cdots, N$) are sampled from the prior probability distribution in the parameter space. When the prior probability distribution is the uniform distribution, the model parameter sets are sampled by the grid search. At each sampled point in the parameter space, particles are placed.

At time $t$, the $n$-th parameter set $\boldsymbol{\theta}_n$, and the $m$-th road segment, the similarity between observed and simulated mean velocities is evaluated using the two error functions below. One is the percentage error $E^p_{t,n,m}(\boldsymbol{\theta}_n)$,

$$E^p_{t,n,m}(\boldsymbol{\theta}_n) \equiv \frac{\left|v^{\text{sim}}_{t,n,m}(\boldsymbol{\theta}_n) - v^{\text{obs}}_{t,m}\right|}{v^{\text{obs}}_{t,m}} \times 100\%, \tag{3}$$

where $v^{\text{sim}}_{t,n,m}(\boldsymbol{\theta}_n)$ is the simulated mean velocity at time $t$ and $m$-th segment by $\boldsymbol{\theta}_n$ and $v^{\text{obs}}_{t,m}$ is the observed one at the same time and segment. The other error function is absolute error $E^a_{t,n,m}(\boldsymbol{\theta}_n)$,

$$E^a_{t,n,m}(\boldsymbol{\theta}_n) \equiv \left|v^{\text{sim}}_{t,n,m}(\boldsymbol{\theta}_n) - v^{\text{obs}}_{t,m}\right|. \tag{4}$$

Using $E^p_{t,n,m}(\boldsymbol{\theta}_n)$ and $E^a_{t,n,m}(\boldsymbol{\theta}_n)$, two likelihood functions are defined.

$$\mathcal{L}_{t,n}^{p}(\boldsymbol{\theta}_n) \equiv \prod_m \frac{1}{\sqrt{2\pi\sigma_p^2}} \exp\left[-\frac{\{E_{t,n,m}^p(\boldsymbol{\theta}_n)\}^2}{2\sigma_p^2}\right], \quad (5)$$

$$\mathcal{L}_{t,n}^{a}(\boldsymbol{\theta}_n) \equiv \prod_m \frac{1}{\sqrt{2\pi\sigma_a^2}} \exp\left[-\frac{\{E_{t,n,m}^a(\boldsymbol{\theta}_n)\}^2}{2\sigma_a^2}\right], \quad (6)$$

where $\sigma_p$ and $\sigma_a$ are hyperparameters that control the tolerance of $E_{t,n,m}^p(\boldsymbol{\theta}_n)$ and $E_{t,n,m}^a(\boldsymbol{\theta}_n)$, respectively. This paper adopts $\sigma_p = 10\%$ and $\sigma_a = 10$ km/h. Product by $m$ in **Equation 3** and **4** imposes higher error than the summation if any of $v_{t,n,m}^{\text{sim}}(\boldsymbol{\theta}_n)$ is largely deviated to $v_{t,m}^{\text{obs}}$. Namely, $\mathcal{L}_{t,n}^p(\boldsymbol{\theta}_n)$ and $\mathcal{L}_{t,n}^a(\boldsymbol{\theta}_n)$ decreases even if any of simulated mean velocities at any segments is not close to observed one.

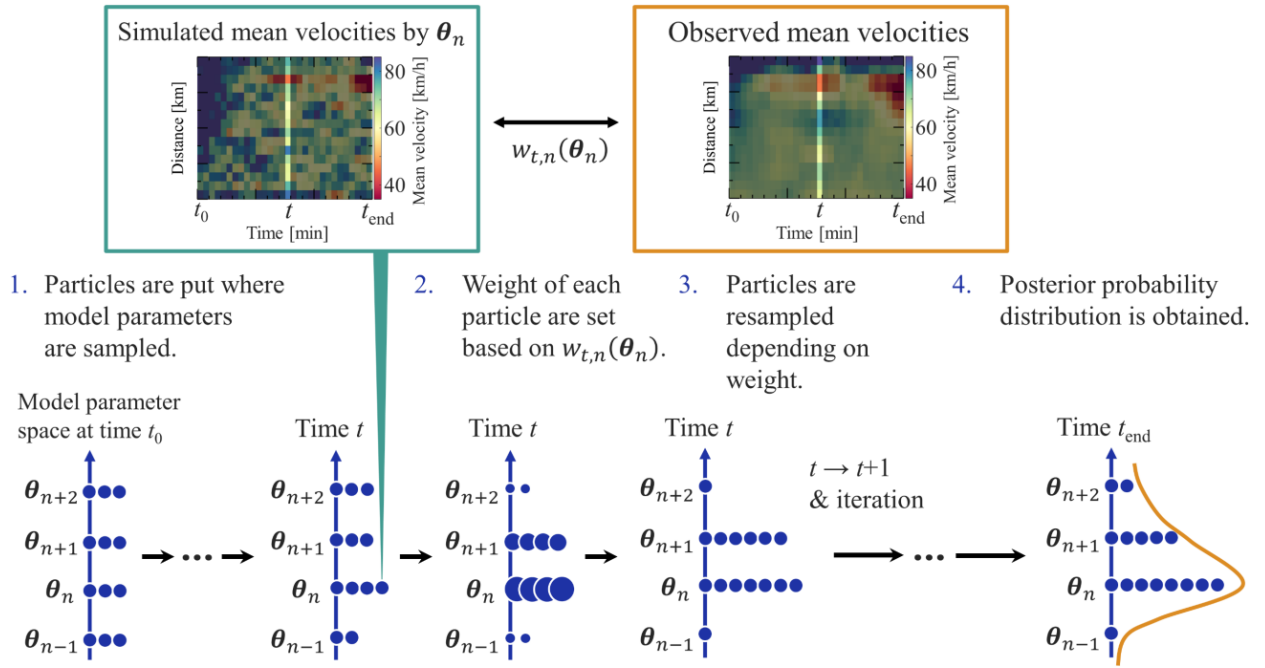

**Figure 4** Schematic view of the optimal model parameter estimation by the particle filter. In each time step, weight is calculated as the similarity of mean velocities between simulated and observed data at time *t*. High weight is expressed as large particles. When the weight is high, particles at the model parameter set increases. After iterating until $t_{\text{end}}$, the posterior probability distribution is obtained. Here, the optimal values are the *n*-th parameter set $\boldsymbol{\theta}_n$.

Using the two likelihoods, the weight of $\boldsymbol{\theta}_n$ is defined as

$$w_{t,n}(\boldsymbol{\theta}_n) \equiv \left[\ln \mathcal{L}_{t,n}^p(\boldsymbol{\theta}_n) + \ln \mathcal{L}_{t,n}^a(\boldsymbol{\theta}_n)\right]^{-2}. \quad (7)$$

The definition of the weight is determined for the reasons below. The percentage error imposes larger error in a low velocity range. In contrast, the absolute error imposes an equal error in all velocity range. Thus, $\ln \mathcal{L}_{t,n}^p(\boldsymbol{\theta}_n)$ is high when congestion is reproduced in the simulation even though reproduced congestion is overestimated. As a result, the weight of the model parameters which reproduce severe congestion tends to high when using only $\ln \mathcal{L}_{t,n}^p(\boldsymbol{\theta}_n)$ in the definition of the weight. On the other hand, $\ln \mathcal{L}_{t,n}^a(\boldsymbol{\theta}_n)$ is high when major mean velocities in observed data are reproduced. Therefore, in the early stage of congestion, the weight tends to be high at the model parameters which reproduce free flow. Consequently, congestion cannot be simulated with the estimated model parameters optima. Therefore, $w_{t,n}(\boldsymbol{\theta}_n)$ based on both the percentage and absolute error defined in **Equation 7** can estimate appropriate model parameters independent of observed mean velocities. Since $\mathcal{L}^p_{t,n}(\boldsymbol{\theta}_n)$ and $\mathcal{L}^a_{t,n}(\boldsymbol{\theta}_n)$ are always less than unity because of the product of the probability densities, $\ln \mathcal{L}^p_{t,n}(\boldsymbol{\theta}_n) + \ln \mathcal{L}^a_{t,n}(\boldsymbol{\theta}_n)$ is negative. Thus, $w_{t,n}(\boldsymbol{\theta}_n)$ is always positive and the higher $\ln \mathcal{L}^p_{t,n}(\boldsymbol{\theta}_n) + \ln \mathcal{L}^a_{t,n}(\boldsymbol{\theta}_n)$ returns the higher $w_{t,n}(\boldsymbol{\theta}_n)$. Both percentage error and absolute error are high, the weight is high, which indicates $\boldsymbol{\theta}_n$ is a good parameter set to reproduce the observed mean velocities. After $w_{t,n}(\boldsymbol{\theta}_n)$ is obtained for all *N* parameter sets, each weight is normalized,

$$w_{t,n}(\boldsymbol{\theta}_n) \leftarrow \frac{w_{t,n}(\boldsymbol{\theta}_n)}{\sum_{n=1}^{N} w_{t,n}(\boldsymbol{\theta}_n)}. \tag{8}$$

Particles at $\boldsymbol{\theta}_n$ are resampled according to $w_{t,n}(\boldsymbol{\theta}_n)$. As a result, the number of particles at $\boldsymbol{\theta}_n$ that reproduce observed mean velocities well increases. Then, moving to the next time *t*+1 and the same processes are applied. PPDs of each model parameter are obtained from histogram of $w_{t,n}(\boldsymbol{\theta}_n)$ in the final time step $t_{\text{end}}$. Optimal values of model parameters are identified from the maximum a posteriori (MAP) in this paper.

**Simulation of Traffic Flow for Prediction**
The simulation of traffic flow for prediction is carried out as follows. The initial condition is estimated as the microscopic traffic state from the latest observed mean velocities as shown in **Figure 2** step 3. In this process, $v_f$ and $k_m$ in **Equation 1** are derived from the simulated mean velocities and densities by the optimal model parameters set. The inflow rate until the prediction horizon is also necessary for the simulation. This paper estimates the future inflow rate by extrapolating the trend in traffic counter data because a significant variation in inflow rate is not expected for the short term of ~30 minutes. Therefore, we assume that the result of the estimated inflow rate does not significantly impact the prediction results as compared with the estimation of the optimum model parameters. By adopting the initial condition, the model parameter set optima, and the estimated inflow rate in the future, traffic simulation is carried out as step 4 in **Figure 2**.

**CONGESTION DATA FOR VALIDATION AND SIMULATION SETTINGS**
Our proposed traffic prediction method is validated using two traffic congestion scenarios observed on two Japanese expressways. One is observed in Ken-O Expressway, which is the outer-most ring expressway around the Tokyo metropolitan area. The other is Odawara-Atsugi Road, which is a radial expressway from Tokyo to a suburb. **Figure 5** shows spatiotemporal mean velocities in these congestion scenarios obtained from DFOS. The two scenarios show a slowly and rapidly evolving congestion. The former scenario is useful to validate our proposed method for ordinal congestion. Whereas, in the latter scenario, the real-time performance of the prediction is more important than the former. Hence, these two congestion scenarios are suitable for evaluating the scalability of our proposed method.

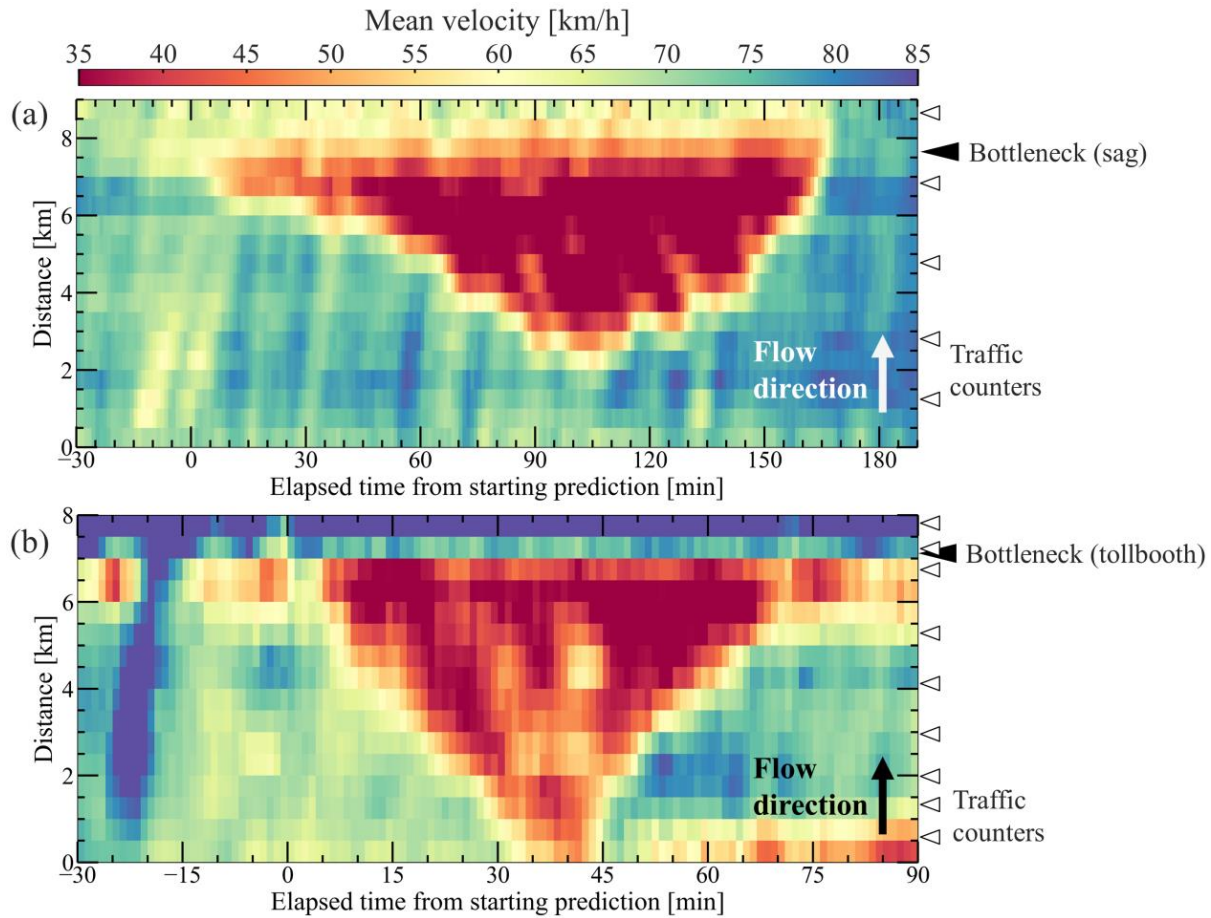

**Figure 5 Observed mean velocities in 500-meter, 1-minute patches obtained by DFOS on (a) Ken-O Expressway, and (b) Odawara-Atsugi Road. The position of the bottleneck and traffic counters are also shown in the black and white arrows on the right edge.**

    In Ken-O Expressway, the road length for the validation is 9 km where the sag section is present around the 7.6 km position. Due to this, the congestion occurred and lasted for approximately 2 hours. This paper defines a congestion as a road segment with mean velocity below 40 km/h, which is a common definition in Japan. **Figure 6 (a)** also shows traffic volume as the inflow rate obtained by the traffic counter.

    In Odawara-Atsugi Road, the road length for the validation is 8 km. At the 7.1 km location, there is a tollbooth as the bottleneck. It causes rapidly growing and dissipating congestion in contrast to that on Ken-O Expressway. The congestion lasts only about 1 hour. Time series data of traffic volume is also shown in **Figure 6 (b)**.

    Simulation settings are as follows. These expressways are both 2-lane roads under the speed limit of 80 km/h. Thus, the road model consists of two lanes, and the speed limit of the slow and fast lane is set as 80 and 100 km/h, respectively. The CA model configuration and the sampling of model parameters are listed in **Table 1**. In the bottleneck sections, this paper introduces different random braking probability $p^{BN}$, which is higher than that outside the bottleneck, to control the strength of the bottleneck. Outside the bottleneck, random braking probability $p$ is adopted. The sampling ranges and step sizes are determined by test simulations. The number of each sampled parameter of $p^{BN}$, $p$, $q$, and $r$ is 13, 5, 5, and 5, respectively. Therefore, the total number of model parameter sets is $13 \times 5 \times 5 \times 5 = 1625$. From the 1625 simulation data sets by these parameter sets and the observed mean velocities for the last 30 minutes, PPDs of the

model parameters $\boldsymbol{\theta} = (p^{BN}, p, q, r)$ are derived and their optimal values that reproduce observed mean velocities are estimated.

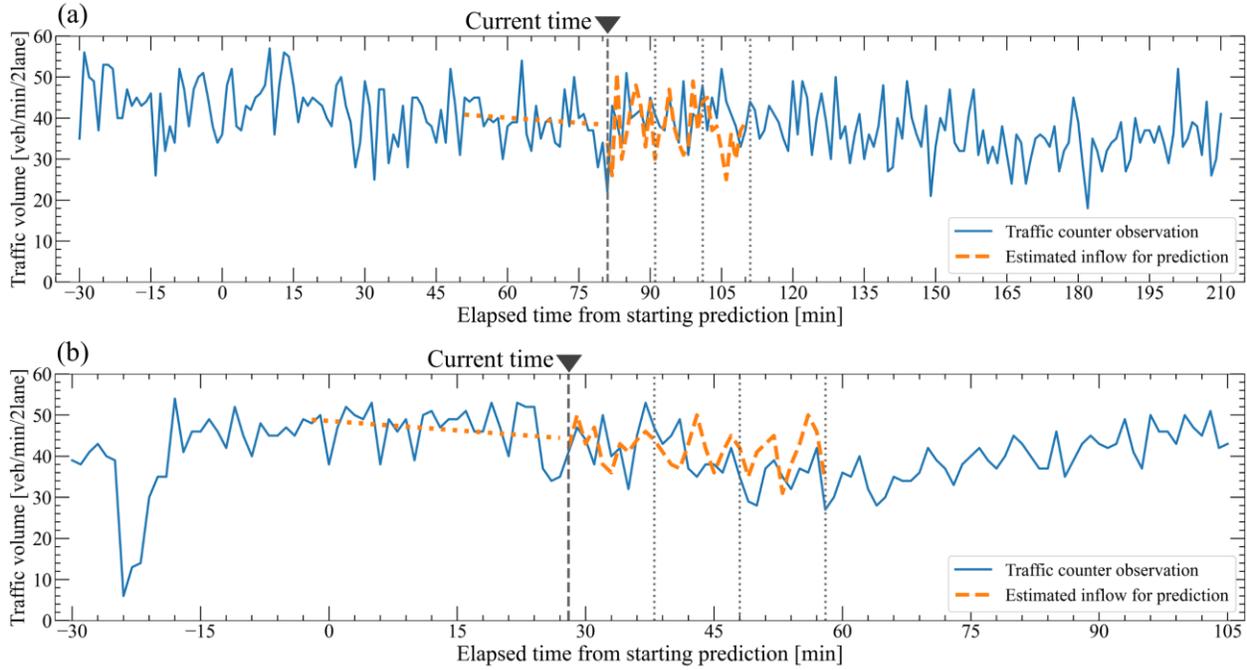

**Figure 6** Traffic volume in (a) Ken-O Expressway and (b) Odawara-Atsugi Road used for validation observed by the traffic counter. As examples, estimated future inflow rate at (a) $t = 81$ and (b) $t = 28$ minutes for the prediction simulation is also denoted as examples in orange with the linear regression for the past 30 minutes. Vertical dotted lines are the time in the next 10, 20, and 30 minutes from the current time.

**TABLE 1 Simulation settings**

| Cell-Automata Configuration | Value | | |
|---|---|---|---|
| Cell length | 10 m | | |
| Time step size | 1.8 sec | | |
| Corresponding velocity resolution | 20 km/h | | |
| Bottleneck position (Ken-O Expressway) | 7.6–7.8 km | | |
| Bottleneck position (Odawara-Atsugi Road) | 6.9–7.1 km | | |
| | | | |
| **S-NFS Model Parameters** | **Lower Limit** | **Upper Limit** | **Step Size** |
| Bottleneck random braking effect prob. $p^{BN}$ | 0.26 | 0.50 | 0.02 |
| Random braking effect prob. $p$ | 0.05 | 0.25 | 0.05 |
| Slow-to-start effect prob. $q$ | 0.05 | 0.25 | 0.05 |
| Anticipation prob. $r$ | 0.91 | 0.99 | 0.02 |

In the simulation for the prediction after the optimal model parameters are estimated, the initial condition estimated from the latest observed mean velocities are applied. The fraction of vehicles on the fast lane is assumed to be 60%, which is the typical value for Japanese expressways in the high occupancy case. Besides, it is necessary to estimate the future inflow rate for the prediction simulation. It is estimated from the linear trend and random noise based on the measurement data of the traffic counter located at the

Y. Yajima, H. Prasad, D. Ikefuji, T. Suzuki, S. Tominaga, H. Sakurai, and M. Otani

closest position corresponding to the origin of the road model. The linear trend and standard deviation are derived from the data for the last 30 minutes. The orange dotted and dashed lines in **Figure 6** show the examples.

The traffic simulation considers the lane changing process, where in each time step, vehicles check the velocity in the lane-changed and not-changed case. Suppose a vehicle achieves higher velocity if it changes the lane. In that case, if the velocity difference and distance to the nearest rear vehicle are sufficient to avoid collision, the vehicle changes the lane with a probability. The probability of lane changing is set as 10%. With this setting, the probability that a vehicle changes the lane at least once within 33 time steps (59.4 seconds ~ 1 minute) is 96.9% when all conditions of the lane changing are satisfied during the period.

**RESULTS AND DISCUSSION**

**Prediction of Mean Velocity**
**Figure 7** shows mean absolute error (MAE), $E_{\text{MAE}}$, of predicted mean velocities averaged over 500-meter, 1-minute spatiotemporal patches in Ken-O Expressway and Odawara-Atsugi Road. The prediction horizon is 10, 20, and 30 minutes. The MAE of prediction horizon $T_h$ is defined by **Equation 9**,

$$E_{\text{MAE}} = \frac{1}{MT_h} \sum_{m=1}^{M} \sum_{t=T_s}^{T_s+T_h} |v_{t,m}^p - v_{t,m}^g|, \quad (9)$$

where $T_s$ is the prediction start time, $v_{t,m}^p$ and $v_{t,m}^g$ are predicted and ground-truth mean velocity at a spatiotemporal patch at time *t* and the *m*-th road segment, respectively. It attains below 15 km/h through most of the period on both expressways. **Figure 8** and **9** also show observed, predicted, and ground-truth mean velocities and derived PPDs of model parameters at several points in time denoted in (a1)–(b7) in **Figure 7**.

The MAEs show a trend depending on the time-evolution of congestion. For example, the MAEs increase prior to the observation of congestion due to underestimated congestion in the prediction as shown in **Figure 8** (a1) and **Figure 9** (b1). This is because the optimal value of $p^{\text{BN}}$ is not high enough to reproduce congestion. Since the observed traffic flow does not show congestion yet, the model parameter which does not reproduce congestion is accepted. Once congestion is observed, the results correctly predict growing congestion as shown in **Figure 8** (a2), (a3), and **Figure 9** (b2) because the optimal value of $p^{\text{BN}}$ shifts higher, which corresponds to the strong bottleneck. At such times, the MAEs decrease to ~10 km/h even for the prediction horizon of 30 minutes.

In addition, deviation of estimated inflow rate in the future decreases the prediction performance, which is contrary to our original assumption. For example, **Figure 8** (a4) shows that predicted congestion in 20–30 minutes is overestimated whereas the congestion is gradually dissipating in the ground truth. This is due to overestimated inflow rate. Hence, the MAE for the prediction horizon of 30 minutes in **Figure 7** is high at the time (a4). Estimated inflow rate also affects prediction results in the next 20–30 minutes when congestion starts to dissipate as shown in **Figure 9** (b3). In this case, inflow rate decreases in approximately 20 minutes, however, the current method cannot estimate this trend. As a result, the MAEs, especially in the prediction horizon of 30 minutes, increase again before the turning point of the congestion evolution. When the deviation of the estimated inflow returns to small, the MAEs of predicted mean velocity decrease like the time point of (a5) in **Figure 7** and **8**. In contrast, the dissipation of congestion can be predicted beforehand even in the prediction horizon of 30 minutes if inflow rate is correctly estimated. **Figure 9** (b4) shows a clear dissipation in the prediction because of the correct estimation of inflow rate as well as the model parameter estimation.

Congestion dissipation is not clear in **Figure 9** (b5) although a decreasing inflow rate is estimated. This is due to the underestimated anticipation probability $r$. When $r$ decreases, the traffic capacity of the road and the critical density decreases (see (*23*)). Results of **Figure 9** (b4) and (b5) indicate that it will be possible to immediately predict dissipation of congestion when decreasing inflow rate and optimal model parameters are correctly estimated.

When dissipating congestion begins to be observed, the tendency of congestion can be well predicted as shown in **Figure 8** (a6) and **Figure 9** (b6). The bottleneck strength parameter $p^{BN}$ also returns to the lower value. The disappeared time of congestion can also be predicted within an error of a few minutes like **Figure 8** (a7) and **Figure 9** (b7). The MAEs are around 10 km/h or below at such times of the dissipation congestion stage.

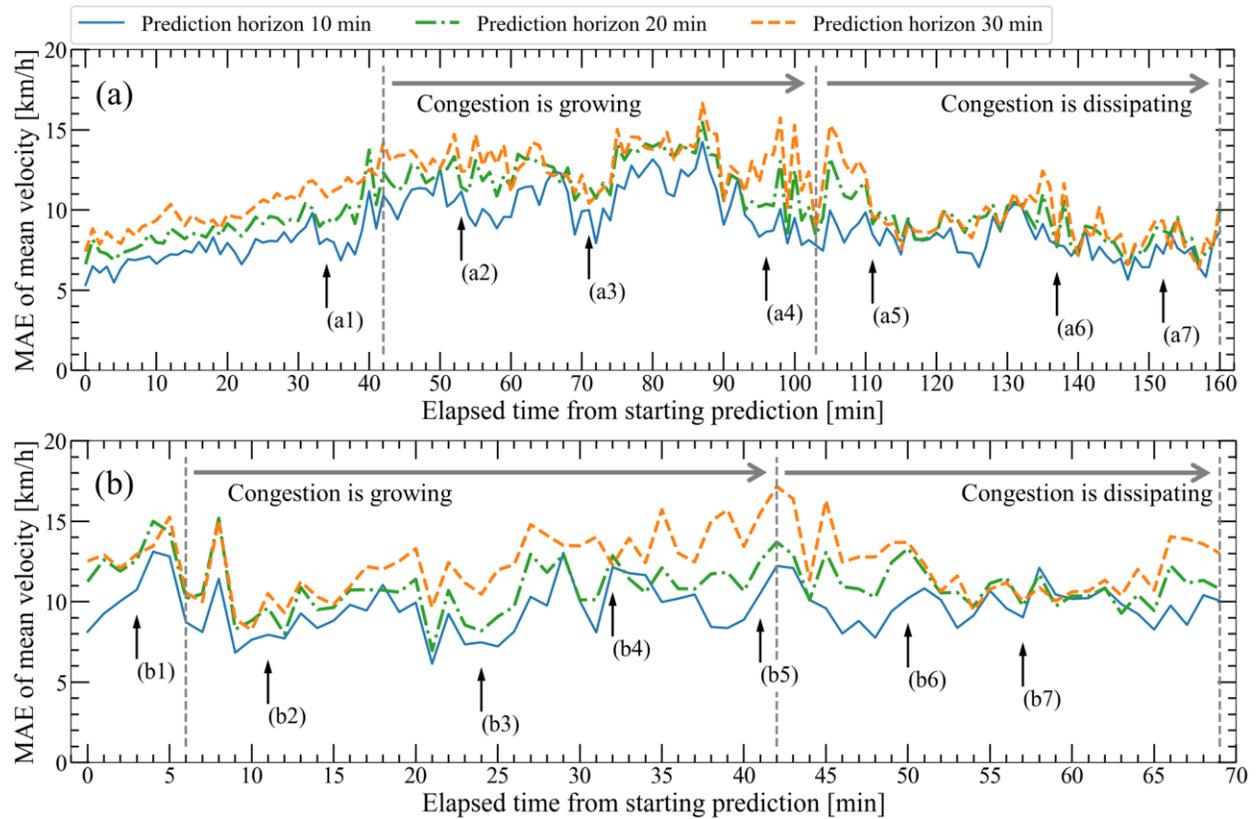

**Figure 7 Mean absolute error (MAE) of predicted mean velocities as a function of elapsed time on (a) Ken-O Expressway and (b) Odawara-Atsugi Road averaged over a 500-meter, 1-minute patch. Vertical dashed lines indicate the times when congestion starts, turns into dissipation, and disappears, respectively. Mean velocities at (a1)–(b7) are shown in Figure 8 and 9 as heat maps.**

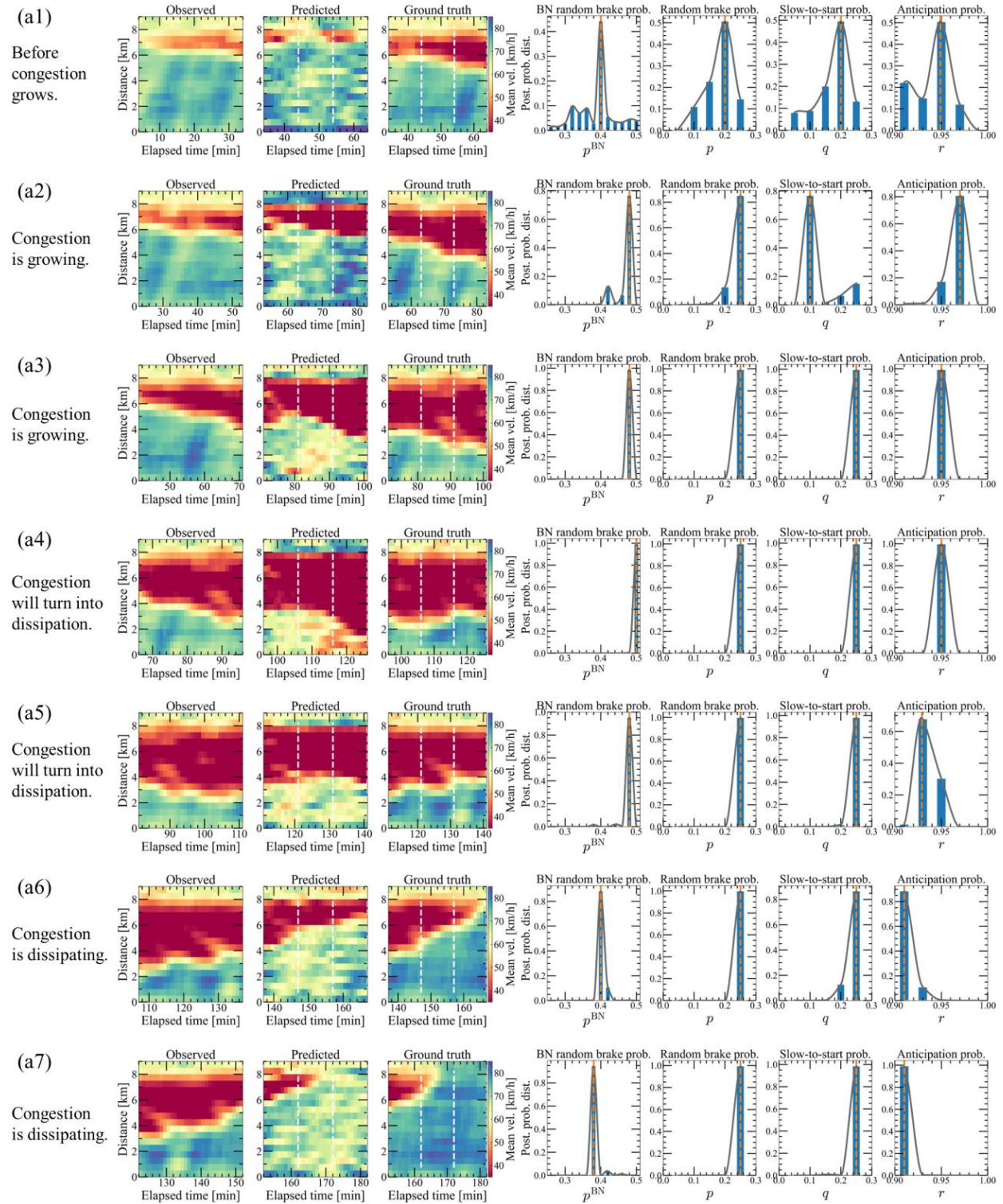

**Figure 8 (From left to right in each row) Observed, predicted, and ground-truth mean velocities, and PPDs of the model parameters for Ken-O Expressway at (a1)–(a7) in Figure 7. The blue bars in the PPDs panels are the results of the particle filter. To derive the MAP, the PPDs are interpolated by the Akima interpolation, a kind of the spline interpolation. The adopted MAPs are shown in vertical orange lines.**

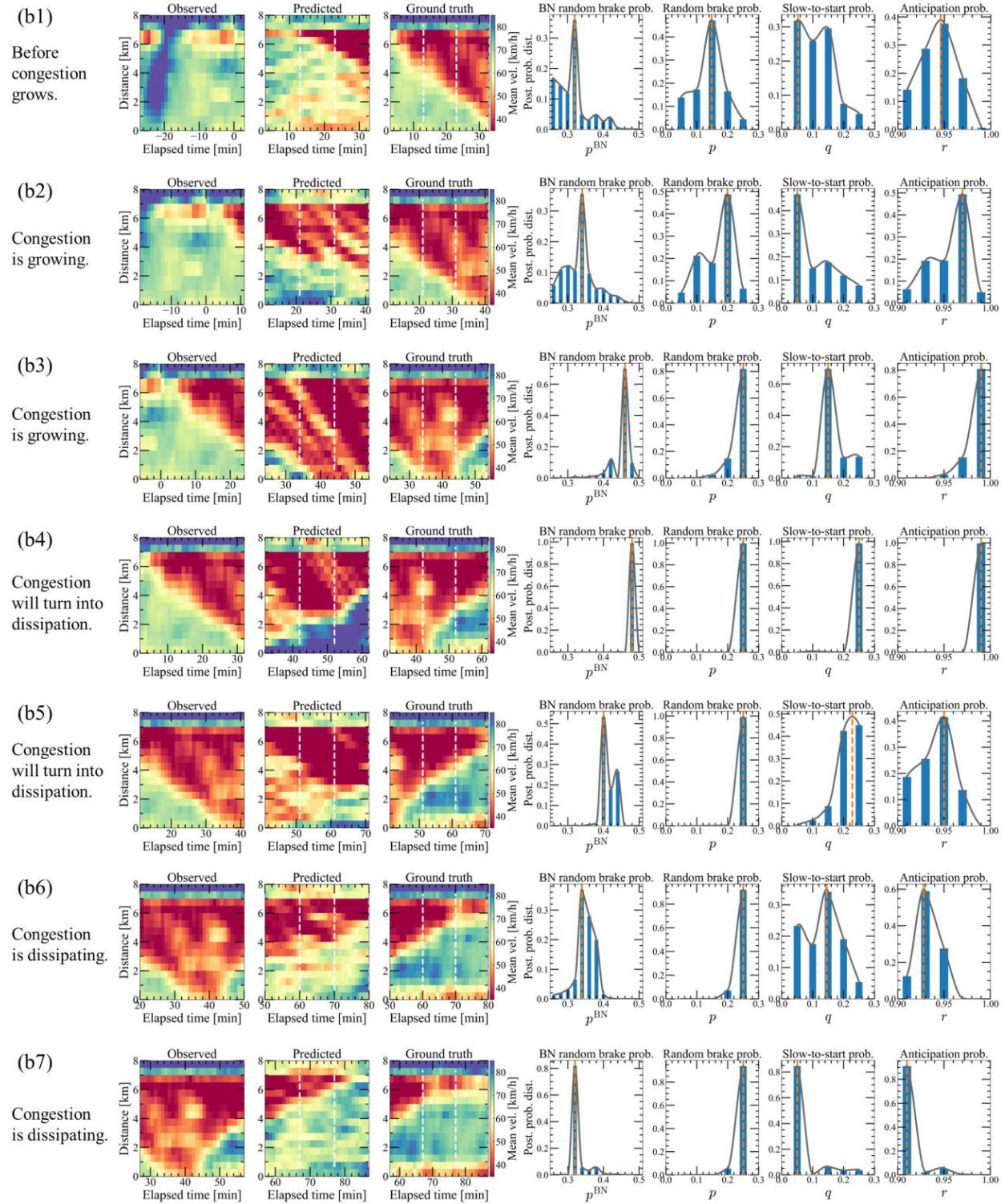

**Figure 9** Same as Figure 8 but for Odawara-Atsugi Road at time (b1)–(b7) denoted in Figure 7.

*Y. Yajima, H. Prasad, D. Ikefuji, T. Suzuki, S. Tominaga, H. Sakurai, and M. Otani*

**Prediction of Congestion Length**
To assess the performance of the proposed method compared with that of conventional sensors, the predicted congestion length is investigated. The congestion length is defined as the total road-segment length where mean velocity is less than 40 km/h. The ground truth is derived from observed DFOS data shown in **Figure 5**. As a conventional sensor, traffic counters are used. The congestion length using traffic counters is predicted from the shock-wave velocity derived from the Lighthill-Whitham-Richards (LWR) model (*33*, *34*), one of the density-wave theory. Namely, the congestion length is predicted as the distance from the bottleneck and the position of the shock front where mean velocity is 40 km/h. **Figure 10** shows the MAEs of predicted congestion length derived from our prediction results and the traffic counters.

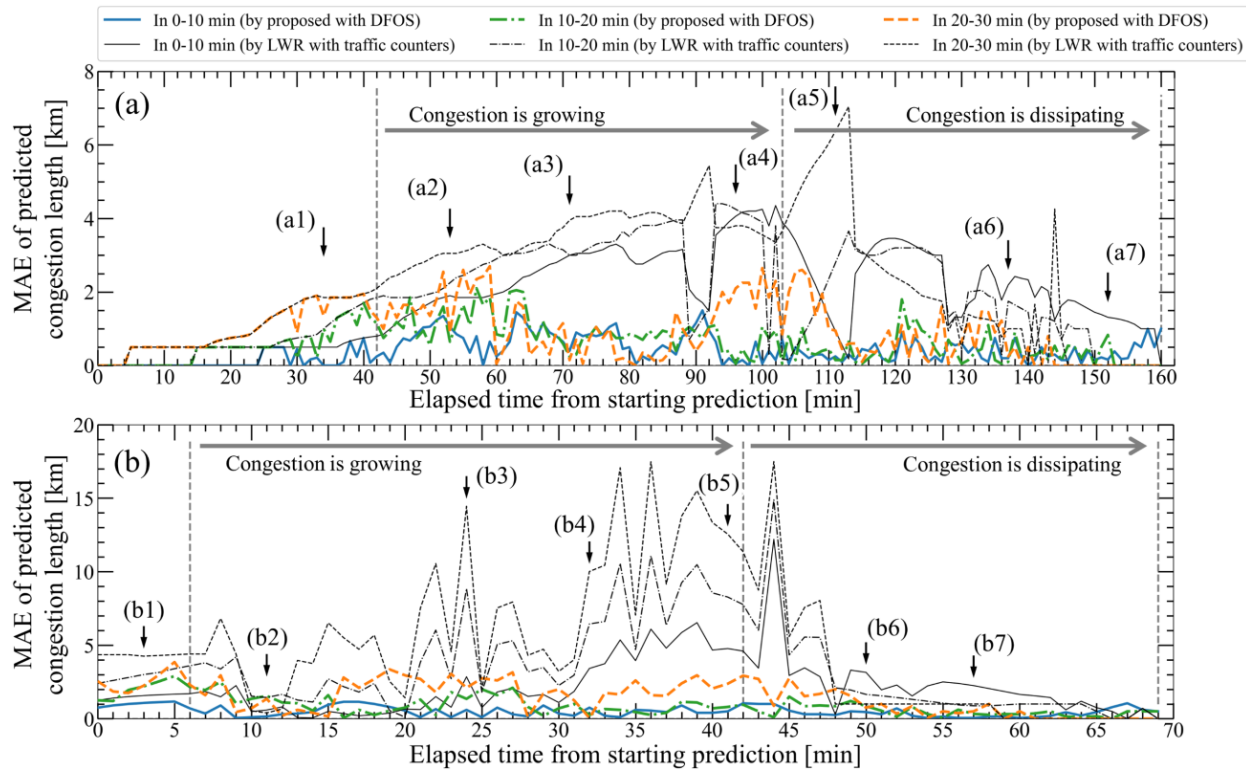

**Figure 10 MAE of predicted congestion-length in 0–10, 10–20, and 20–30 minutes derived from the proposed and conventional method for (a) Ken-O Expressway and (b) Odawara-Atsugi Road. The vertical dashed lines and the labels (a1)–(b7) are the same as Figure 7.**

Figure 10 (a) shows that the MAEs gradually increases in both results of DFOS and point sensors before congestion is observed because it cannot be predicted. However, the MAEs of the proposed method decreases below 1 km after congestion is observed whereas the prediction result by the conventional continues to deviate. This is because there is a time lag to detect congestion by the traffic counters due to slow growing speed of congestion. In the congestion dissipation stage, the same trend of lower MAE by the proposed method is seen. On average, the MAEs of the predicted congestion length in the growing congestion stage ($t = 42–103$ minutes) in the next 0–10, 10–20, and 20–30 minutes decreases by 75%, 68%, and 67%, respectively when using the proposed method. In the dissipating congestion stage ($t > 103$ minutes), the MAEs in 0–10, 10–20, and 20–30 minutes decreases by 83%, 76%, and 69%, respectively.

Figure 10 (b) shows the significant improvement in prediction accuracy for the proposed method whereas the conventional method cannot achieve the accuracy despite the higher density of the traffic counters on this expressway. Since the growing and dissipation speed of congestion on the expressway, the traffic counters cannot properly track the shock-wave velocity. In contrast, DFOS can follow the current congestion and predict the time-evolution of congestion. On average, the MAEs of predicted congestion length through the growing-congestion stage ($t = 6$–43 minutes) in the next 0–10, 10–20, and 20–30 minutes decreases by 72%, 73%, and 71%, respectively when using the proposed method. The MAEs in the dissipating-congestion stage ($t > 43$ minutes) decreases by 82%, 79%, and 69% for the same prediction horizons. Even in the case of rapidly varying congestion case, our proposed method with DFOS attains much lower error than the conventional sensor and method.

**Prediction of Travel Time**

In addition to the congestion length, travel-time prediction is investigated. The ground-truth travel time is derived from observed mean velocities by DFOS (**Figure 5**). For the comparison, the instantaneous travel time (ITT) derived from the traffic counters, which is widely used as a conventional method, is applied. **Figure 11** shows error of predicted travel time to reach the terminal of the road.

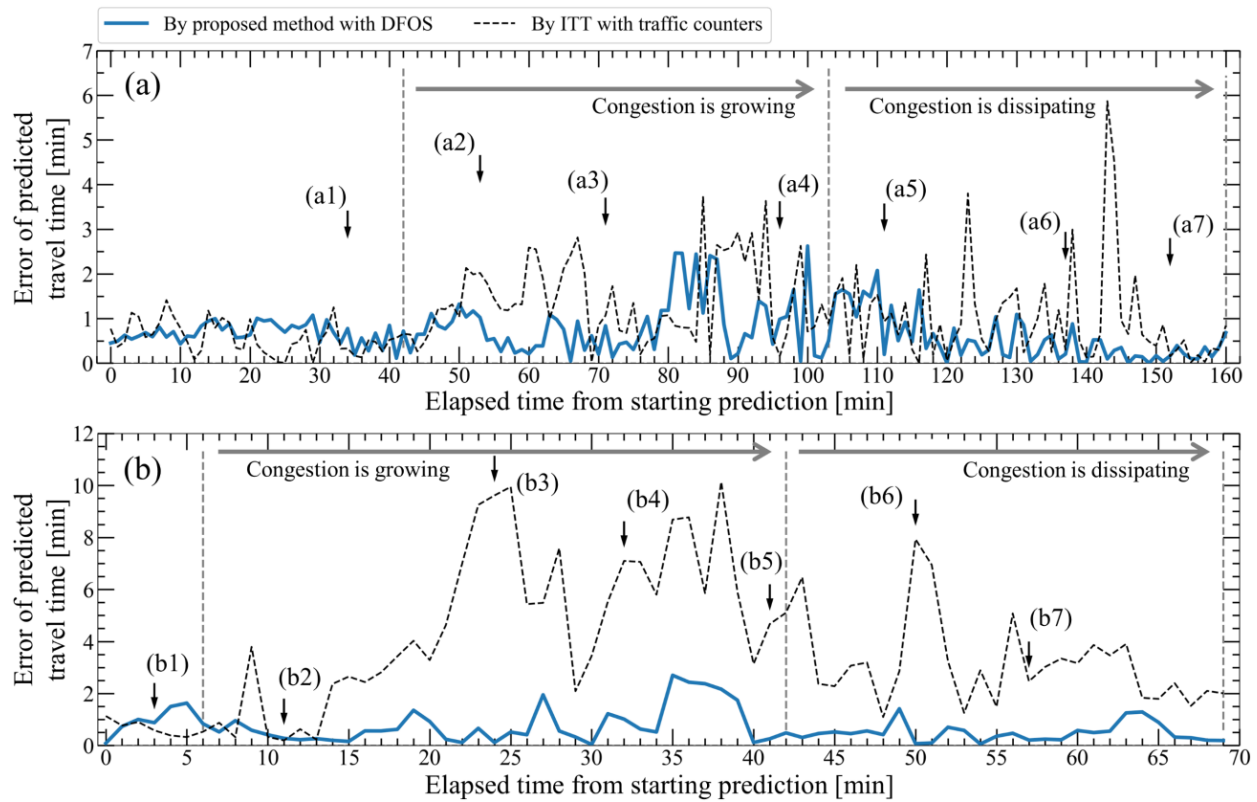

**Figure 11 Error of travel-time prediction derived from the prediction result by DFOS and the instantaneous travel time (ITT) from traffic counter data on (a) Ken-O Expressway and (b) Odawara-Atsugi Road.**

Since the growing and dissipation speed of congestion is slow in Ken-O Expressway, **Figure 11** (a) does not show significant improvement between the proposed method and the conventional ITT with the traffic counters. Nevertheless, the MAEs of the proposed method with DFOS attain lower value in the early stage of growing congestion ($t = 50$–70 minutes) than those of the conventional method owing to the real-time data assimilation. The MAE of the ITT becomes high due to the time lag of the congestion observation by the traffic counters. On average, the MAE of travel time in the growing and dissipating congestion stage decreases by 40% and 48%, respectively in the proposed method.

**Figure 11** (b) shows that the prediction error of travel time is clearly suppressed compared to the ITT. In the early stage of congestion ($t < 13$ minutes), both prediction error by our and conventional result is low because congestion is too short to affect travel time. After $t = 13$ minutes, the effect of growing congestion becomes significant. That is, DFOS prediction result successfully consider growing congestion vehicles will experience in the future, however, ITT considers the current short congestion. It leads to underestimated travel time. After congestion starts to dissipate from $t = 42$ minutes, the prediction result by DFOS considers dissipating congestion and attain low error whereas the ITT by traffic counters only considers the current long congestion, which lead to overestimated travel time. On average, travel time error in the growing-congestion stage and dissipating stage decreases by 80% and 84%, respectively. The real-time data assimilation using continuous and dense traffic data by DFOS significantly increases the accuracy of travel time prediction rather than the conventional method.

In summary, mean velocities can be predicted with the MAE around 10–15 km/h even when the prediction horizon is 30 minutes, through the congestion period. These accurate predictions of mean velocities make it possible to predict more correct congestion length and travel time. In particular, the rapid growing and dissipating congestion case shows significant improvement in the prediction. This is achieved with the proper valid data assimilation based on the proposed method in real-time and continuous traffic data measured by DFOS, which enables the reliable and accurate simulation. The obtained results can be useful for road managers and operators in practical traffic scenarios to understand traffic congestion properties by predicting time for growing and dissipation, its length, and travel time to provide information to drivers.

Although our proposed prediction method attains better performance compared with the conventional methods, there is room for further improvement toward more accurate prediction than the current achievement. The performance can be firstly improved by predicting occurrence of congestion prior to the observation of congestion in DFOS data. This will be achieved by taking account of the low posterior probability at high $p^{\mathrm{BN}}$ as shown in **Figure 8** (a1) and **Figure 9** (b1) for the prediction. Secondly, adopting sophisticated models to estimate inflow rate used for the prediction simulation is considered instead of the simple extrapolation of the trend as shown in **Figure 6**. They are future prospects of this work.

**CONCLUSIONS**

We have demonstrated a real-time short-term traffic flow prediction based on data assimilation and DFOS. The S-NFS model is adopted as the theoretical model for data assimilation. The optimal values of the model parameters in the S-NFS are estimated from observed and simulated mean velocities through the particle filter. The proposed method is validated using two congestion scenarios which have different evolution speeds. The main results of this paper are as follows.

1). The MAEs of predicted mean velocities mainly range 10–15 km/h even in the prediction horizon of 30 minutes. Although the MAEs increase when congestion is not yet observed in DFOS data, they decrease below 10 km/h once congestion is observed. It is also possible to predict that congestion turns into dissipation in some cases before it is observed in DFOS data.
2). The error in predicted congestion length is significantly improved from the conventional method and the traffic counters. The prediction error of congestion length decreases by ~70–80% in both congestion scenarios.
3). The error in travel-time prediction also improves in our proposed method as compared with the conventional method, especially in the rapidly growing and dissipating congestion scenario where the real-time prediction is more important. The prediction error decreases by ~80% in this scenario and ~40% in the slowly growing and dissipating congestion scenario.

Owing to the spatially continuous and real-time properties of DFOS, data assimilation using traffic data of DFOS enables immediate traffic flow prediction with much higher accuracy than the conventional method. Moreover, since optical fibers are usually installed for telecommunication along expressways, our prediction method is promising to be applied to other roads. This study suggests that the proposed method can realize traffic flow optimization and maximize traffic capacity toward the intelligent transportation system.

**AUTHOR CONTRIBUTIONS**
The authors confirm contribution to the paper as follows: study conception and design: Y. Yajima, H. Prasad, D. Ikefuji, T. Suzuki, S. Tominaga, H. Sakurai; data collection: D. Ikefuji, T. Suzuki, S. Tominaga, M. Otani; analysis and interpretation of results: Y. Yajima, H. Prasad, D. Ikefuji; draft manuscript preparation: Y. Yajima, H. Prasad, D. Ikefuji, T. Suzuki, S. Tominaga, H. Sakurai, M. Otani. All authors reviewed the results and approved the final version of the manuscript.